# Hebbian Crosstalk Prevents Nonlinear Unsupervised Learning

Kingsley J.A. Cox and Paul R. Adams

Department of Neurobiology, SUNY Stony Brook, NY, USA
Kalypso Institute, Stony Brook, NY, USA

**Abstract**

Learning is thought to occur by localized, experience-induced changes in the strength of synaptic connections between neurons. Recent work has shown that activity-dependent changes at one connection can affect changes at others ("crosstalk"). We studied the role of such crosstalk in nonlinear Hebbian learning using a neural network implementation of Independent Components Analysis (ICA). We find that there is a sudden qualitative change in the performance of the network at a critical crosstalk level and discuss the implications of this for nonlinear learning from higher-order correlations in the neocortex.

**Abbreviations:**
CaM kinase, $Ca^{2+}$/calmodulin-dependent protein kinase; LTP, long-term potentiation; LTD, long-term depression; ICA, independent components analysis; NMDAR, N-methyl -D-aspartate receptor

* To whom correspondence should be addressed. E-mail: kcox@notes.sunysb.edu

**Author Summary**

The brain extracts information from the environment, so its owner can better survive and reproduce. This is achieved by individual adjustments of the strengths of the synapses that form connections between neurons, as a result of their ongoing electrical activity. However, recent experimental work suggests that the accuracy of such adjustments, while very high, is not perfect. Key intracellular messengers (such as calcium) can diffuse between synapses, leading to "crosstalk", so changes at one connection depend on changes at others. We propose that the accuracy of these synaptic adjustments must be very high for sophisticated learning of "deep" features of the world, which generate complex "higher-order" patterns of correlated neural activity. Learning from higher-order correlations, for which the neocortex seems specialized, requires nonlinear adjustment rules. In this work we explore the role of synaptic adjustment crosstalk errors, in a simple, popular nonlinear neural network learning paradigm. We find there is a critical crosstalk level above which the network cannot usefully learn. This crosstalk error level is typically comparable to the small but inevitably finite level found in the brain. We suggest that the main task confronting the neocortex, to which it devotes most of its circuitry, is crosstalk mitigation. Such crosstalk mitigation would ultimately allow understanding of the world and the emergence of mind.

**Introduction**

Unsupervised artificial neural networks usually use local, activity-dependent (and often Hebbian) learning rules, to arrive at efficient, and useful, encodings of inputs in a self-organizing manner [1,2]. It is widely believed that the brain, and particularly the neocortex, might self-organize, and efficiently represent an animal's world, in a similar way [3,4], especially since synapses exhibit spike-coincidence-based Hebbian plasticity [5-7], such as long-term potentiation (LTP) or long-term depression (LTD). However, some data [8-12] suggest that biological Hebbian learning may not be completely synapse-specific, and other data [13,14], while showing a high degree of specificity, do not unequivocally show complete specificity. Very recent data [15] has shown that induction of LTP at one synapse modifies the inducibility of LTP at closely neighboring synapses ("crosstalk"). Perhaps, given the close packing of synapses in neuropil (> $10^9$ $mm^{-3}$ in neocortex; [16]), complete chemical isolation may be impossible. Such crosstalk, although typically very small, would nevertheless be a possible source of synaptic adjustment inaccuracy.

Biological processes are often noisy and inaccurate, and slight Hebbian inspecificity might not matter. However, there is one biological process that exhibits truly extraordinary specificity, the copying of DNA. Genomes "learn" from their environment by Darwinian evolution, which requires extraordinarily accurate self-replication [17-20]; brains learn from their environment by Hebbian adjustments of individual synapses, and we suspect that this form of biological learning also requires extraordinary accuracy in the elemental step, in this case spike-timing-dependent synaptic strengthening. Thus synaptic update errors, although small, may have important consequences for sophisticated, particularly nonlinear, learning

To investigate this, we investigated update inspecificity in a neural network model of a simple but powerful approach, independent components analysis (ICA), that has been proposed as a model for nonlinear learning from higher-order correlations in the neocortex [21-25]. With ICA it is possible to recover the unknown irreducible sources that give rise, via a linear but unknown mixing process, to sensory data. The independent components (ICs) of natural scenes resemble the oriented edge detectors found in primary visual cortex [26-28].

Here we describe computer experiments that suggest that slight Hebbian inspecificity, or crosstalk, can destabilize learning in simple ICA networks. We point out possible implications of these results for understanding neocortical circuitry and physiology. In particular, crosstalk increases with synapse density, imposing a low limit on the number of learnable inputs to a single neuron. We therefore propose that the main task for the neocortex is raising this limit.

# Methods

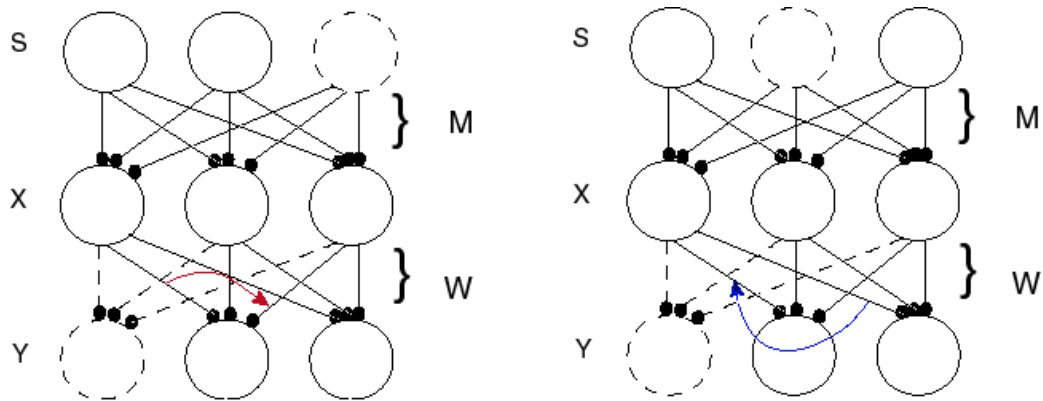

Figure 1 Schematic ICA network. Mixture neurons X receive weighted signals from independent sources S, and output neurons Y receive input from the mixture neurons. The goal is for each output neuron to mimic the activity of one of the sources, by learning a weight matrix **W** that is the inverse of **M**. In the diagrams this is indicated by the source shown as a dotted circle being mimicked by one of the output neurons (dotted circle) with the dotted line connections representing a weight vector which lies parallel to a row of $M^{-1}$ i.e. an independent component or "IC". The effect of synaptic update error is represented by curved colored arrows, red being the postsynaptic case (left diagram), and blue the presynaptic case (right diagram). In the former case part of the update appropriate to the connection from the left X cell to the middle Y cell leaks to the connection from the right X cell to the middle Y cell. In the latter case, part of the update computed at the connection from the left X cell onto the right Y cell leaks onto the connection from the left X cell onto the middle Y cell. However, in both these cases for clarity only 1 of the $n^2$ possible leakage paths that comprise the error matrix **E** are shown. Note that learning of **W** is driven by the activities of X cells (the vector **x**) and by the nonlinearly transformed activities of the Y cells (the vector **y**), as well as by an "antiredundancy" process.

Simulations were done using Matlab. Except for figure 5, all simulations stored data only for every hundredth run, or epoch.

An *n* dimensional vector of independently fluctuating sources **s** obtained from a defined (usually Laplacian) distribution is mixed using a mixing matrix **M** (generated using Matlab's "rand" function to give an *n* by *n* dimensional matrix with elements ranging from {0,1} and sometimes {-1,1}), to generate an *n* dimensional column vector **M s** = **x**, the elements of which are linear combinations of the sources, the elements of **s**. For a given run **M** was held fixed, and the numeric labels of the generating seeds, and sometimes the specific form of **M**, are given in the Results or Supplementary Information (since the result depended idiosyncratically on the precise **M** used).

The aim is to estimate the sources $s_1, s_2, ..s_n$ from the mixes $x_1, x_2, ..x_n$ by applying a linear transformation **W**, represented neurally as the weight matrix between a set of *n* mix neurons whose activity represents **x** and a set of *n* output neurons, whose activity **u** represents estimates of the sources. When $\mathbf{W} = \mathbf{PM}^{-1}$ the (arbitrarily scaled) sources are recovered exactly (**P** is a permutation/scaling matrix which reflects uncertainties in the order and size of the estimated sources). Although neither **M** nor **s** may be known in advance, it is still possible to obtain an estimate of the unmixing matrix, $\mathbf{M}^{-1}$, if the (independent) sources are non-Gaussian, by maximizing the entropy (or, equivalently, nonGaussianity) of the outputs. Maximizing the entropy of the outputs is equivalent to making them as independent as possible. Bell and Sejnowski [22] showed that the following nonlinear Hebbian learning rule could be used to do stochastic gradient ascent in the output entropy, yielding an estimate of $\mathbf{M}^{-1}$,

$$\Delta \mathbf{W} = \gamma([\mathbf{W}^T]^{-1} + f(\mathbf{u})\mathbf{x}^T)$$

where **u** (the vector of activities of output neurons) = **Wx** and $\mathbf{y} = f(\mathbf{u}) = g''(\mathbf{u})/g'(\mathbf{u})$ where $g(s)$ is the source cdf, and $\gamma$ is the learning rate.

Amari et al [29] showed that even if $f \neq g''/g'$, the algorithm still converges (in the small learning rate limit) to $\mathbf{M}^{-1}$ if certain conditions on f and g are respected.

Bell and Sejnowski derived specific forms of the Hebbian part of the update rule assuming various nonlinearities. For the logistic function $f(\mathbf{u}) = (1 + e^{-\mathbf{u}})^{-1}$ their rule, which we will call the BS rule, (for superGaussian sources) is

$$\Delta \mathbf{W} = \gamma([\mathbf{W}^T]^{-1} + (\mathbf{1} - 2\mathbf{y})\mathbf{x}^T) \qquad \text{Eq (1)}$$

where **1** is a vector of ones. Using Laplacian sources the convergence conditions are respected even though the logistic function does not "match" the Laplacian. The first term is an antiredundancy term which forces each output neuron to mimic a different source; the second term is antihebbian (in the superGaussian case), and could be biologically implemented by spike coincidence-detection at synapses comprising the connection. We also tested the "natural gradient" form of the rule [30], where Eq (1) is postmultiplied by $\mathbf{W}^T\mathbf{W}$, with similar results. However, while this removes the matrix inversion step, the remaining step would require implausible nonlocal, and "backpropagating", learning [26]. We find that a one-unit form of ICA [31], which replaces the matrix inversion step by a more plausible normalization step, is also destabilized by error (unpublished results).

Errors were implemented by postmultiplying the Hebbian part of $\Delta \mathbf{W}$ by an error matrix **E** (components $E_{ij}$; see below), which shifted a fraction $E_{ij}$ of the calculated Hebbian update $(\mathbf{1} - 2\mathbf{y})\mathbf{x}^T$ from the jth connection on an output neuron onto the ith connection on that neuron, i.e. postsynaptic error (Figure 1, left).

$$\Delta \mathbf{W} = \gamma([\mathbf{W}^T]^{-1} + [(\mathbf{1} - 2\mathbf{y})\, \mathbf{x}^T]\, \mathbf{E}) \qquad \text{Eq (2)}$$

Premultiplying by **E** would assign error from the ith connection on a given output neuron onto the jth connection on another output neuron made by the same presynaptic neuron (presynaptic error; Figure 1, right). We will analyse this presynaptic case elsewhere.

The Error Matrix
The errors are implemented ("error onto all", see below) using an error matrix **E**

$$\mathbf{E} = \begin{pmatrix} Q & \varepsilon & \varepsilon & . & . & \varepsilon \\ \varepsilon & Q & \varepsilon & \varepsilon & . & \varepsilon \\ \varepsilon & \varepsilon & Q & \varepsilon & . & \varepsilon \\ . & . & . & . & . & . \\ \varepsilon & . & . & \varepsilon & Q & \varepsilon \\ \varepsilon & \varepsilon & . & . & \varepsilon & Q \end{pmatrix}$$

where Q is the fraction of update that goes on the correct connection and $\varepsilon = (1 - Q)/(n-1)$ is the fraction that goes on a wrong connection. The likely physical basis of this "equal error-onto-all" matrix is explained below. We often refer to a "total error" $E$ which is 1-Q.

Error onto all
The proposed physical basis of the lack of Hebbian specificity studied in this paper is intersynapse diffusion, for example of calcium. This problem would reflect fundamental limits that any computing device operating above absolute zero will encounter. In principle intersynapse diffusion will only be significant for synapses that happen to be located close together, and it seems likely that the actual arrangements of synapses in space and along the dendritic tree will be arbitrary (merely reflecting the happenstance of particular axon-dendrite close approaches) and unrelated to the statistical properties of the input. This would reflect the standard connectionist view that synaptic potentials occurring anywhere on the dendrites are "integrated" at the initial segment, and might not hold if important computations are done in nonlinear dendritic domains [32]. Nevertheless, in the present work we made the assumption that all connection strength changes are equally likely to affect any other connection strength – an idea we call "error onto all". The underlying premise is that there should be no arbitrarily privileged connections – that the neural learning device should function as a "tabula rasa" [33,34] – which is inherent in the connectionist approach. We extend the idea that all connections should be approximately electrically equivalent [35] to suggest that they might also be approximately chemically equivalent. This could also be viewed as a "meanfield" assumption, so that "anatomical fluctuations" (detailed synaptic neighborhood relations) get averaged out in the large *n* limit, because connections turn-over [34, 36, 37] and are multisynapse [38]. Because of the error-onto-all assumption, the diagonal elements, and also the off-diagonal elements, of **E** are all equal, and in the case of complete specificity **E** reduces to the identity matrix implicit in conventional treatments of Hebbian learning. The "quality" Q of the learning process (Q = 1 is complete specificity), would depend on the number of inputs $n$, the dendritic (e.g. calcium) diffusion length constant, the spine neck and dendritic lengths, and buffering and pumping parameters. In the simplest case, with a fixed dendritic length, as $n$ increases the synapse linear density increases proportionately, and one expects Q = 1/(1+$nb$) where $b$ is a "per synapse" error rate. This expression can be derived as follows (see also Discussion). Call the number of existing (silent or not) synapses comprising a connection $\alpha$. The total number of synapses on the dendrite, N, is therefore N = $n\alpha$ and the synapse density $\rho$ is $n\alpha$ /L where L is the dendrite length. Define $x$ as the linear dendritic distance between the shaft origins of two spiny synapses. For $x = 0$, assume that the effective calcium concentration in an unstimulated synapse is an "attenuation" fraction $a$ of that in the head of a synapse undergoing LTP, due to outward calcium pumping along 2 spine necks in series. Assume that calcium decays exponentially with distance along the shaft [39, 40] with space constant $\lambda_c$, and that the LTP-induced strength change at a synapse is proportional to calcium. The expected total strengthening at neighboring synapses due to calcium spread from a reference synapse at x = 0 where LTP is induced, as a fraction of that at the reference synapse, assuming that $\lambda_c$ is much smaller than half the dendritic length, is given by

$$2\rho \int_0^{L/2} a \exp(-x/\lambda_c) dx \approx 2a\rho\lambda_c = 2a\lambda_c N / L = nb$$

where $b = 2\alpha a \lambda_c / L$

$b$ (a "per connection error rate") reflects intrinsic physical factors that promote crosstalk (spine-spine attenuation and the product of the per-connection synapse linear density and $\lambda_c$), while $n$ reflects the effect of adding more inputs, which increases synapse "crowding". Notice that silent synapses do not provide a "free lunch" – they increase the error rate. Although incipient synapses [49] (potential synapses that do not yet exist) do not worsen error, the long-term virtual connectivity they provide cannot be immediately exploited. We ignore the possibility that this extra, unwanted, strengthening, due to diffusion of calcium or other factors, will also slightly and correctly strengthen the connection of which the reference synapse is part (i.e. $n$ is quite large). This treatment leads to an error matrix with 1 along the diagonal and $nb/(n-1)$ offdiagonally. In order to convert this to a stochastic matrix (rows and columns sum to one, as in **E** defined above) we multiply by the factor 1/(1+$nb$), giving Q = 1/(1+$nb$). We ignore the scaling factor (1+$nb$) that would be associated with **E**, since it affects all connections equally, and can be incorporated into the learning rate. It's important to note that while $b$ is typically biologically very small (~ $10^{-4}$; see Discussion ), $n$ is typically very large (e.g. 1000 in the cortex).

The offdiagonal elements $E_{i,j}$ are given by $(1-Q)/(n-1)$. In the results we use *b* as the error parameter but specify in the text and figure legends where appropriate the total error $E = 1-Q$.

Orthogonal mixing matrix

Sometimes an orthogonal mixing matrix $M_O$ was used. A random mixing matrix was orthogonalized using an estimate of the inverse of the covariance matrix **C** of a sample of the source vectors that had been mixed using **M**. **M** was then premultiplied by the decorrelating matrix **Z** computed as follows

$Z = (C^{½})^{-1}$

The input vectors x generated using $M_O$ constructed in this way were thus variably "whitened", to an extent that could be set by varying the size of the sample used to estimate **C**. The performance of the network was measured against a new solution matrix $Q^{-1}$, which is approximately orthogonal, and is the inverse of the original mixing matrix **M** premultiplied by **Z**, the decorrelating, or whitening, matrix

$Q = Z M$

In another approach, perturbations from orthogonality were introduced by adding a scaled matrix (**R**) of numbers (drawn randomly from a Gaussian distribution) to the whitening matrix **Z**. The scaling factor (which we call 'perturbation') was used as a variable for making **Q** less orthogonal, as in Figure 6 (see also Text S1, section 1).

**Results**

**BS Rule with 2 neurons and random M**

We looked at the BS rule for the simplest case where $n = 2$, with a random mixing matrix. Figure 2 shows the dynamics of initial, error-free convergence for each of the 2 weight vectors, together with the behaviour of the system when error is applied. "Convergence" was interpreted as the maintained approach to 1 of one of the cosines of the angles between the particular weight vector and each of the possible rows of $M^{-1}$ (of course with a fixed learning rate exact convergence is impossible; in Figure 2 $\gamma = 0.01$, which provided excellent initial convergence). Small amounts of error, ($b = 0.005$, equivalent to total error $E = 0.0099$, applied at 200,000 epochs) only degraded the performance slightly. However, at a error rate of 0.02 ($E = 0.0384$), which is above a threshold error rate ($b_c = 0.01037$, $E = 0.0203$ see Figure 4A) each weight vector began to undergo periodic spike-like oscillations, which became more rapid at $b = 0.05$ (Figure 2) and even more so at $b = 0.1$ ($E = 0.166$). Figure 2D shows that the individual weights on one of the output neurons smoothly adjust from their correct values when a small amount of error is applied, and then start to oscillate almost sinusoidally when error is increased further. Note that at the maximal recovery from the spike-like oscillations the weight vector does briefly lie parallel to one of the rows of $\mathbf{M}^{-1}$; one could therefore describe the behavior as switching between assignments, though spending most of its time at nonparallel states. Similar behavior was seen with different initializations of $\mathbf{W}$ or $\mathbf{s}$.

**Orbits**

Figure 3 shows plots of the components of both weight vectors (i.e. the 2 rows of the weight matrix, shown in red or blue) against each other as they vary over time. The weight trajectories are shown as error is increased from zero to a subthreshold value and then to increasingly suprathreshold values. The weights first move rapidly from their initial random values to a tight region of weight space (see blow-up in right plot) that corresponds to a choice of correct ICs, where they hover for the first million epochs. The initial IC found is typically the one corresponding to the longest row of $\mathbf{M}^{-1}$, and the weight vector that moves to this IC is the one that is initially closest to it (a repeat simulation is shown in Text S1, section 2; the initial weights were different and so was the choice). Introduction of subthreshold error produces a slight shift to an adjacent stable region of weight space. Introduction of suprathreshold error initiates a new shift into a limit cycle-like orbit. Further increases in error generate longer orbits. The red and blue orbits superimpose, presumably because the 2 weight vectors are equivalent, but the columns of $\mathbf{W}$ are phase-shifted (see orbits shown in Text S1, section 2). In Figure 3 the weights spend roughly equal amounts of time everywhere along the orbits, but at error rates just exceeding the threshold the weights tarry mostly very close to the stable regions seen at just subthreshold error (i.e. the degraded ICs; see Text S1, section 2).

**Varying Parameters**

Figure 4A summarises results for a greater range of error values using the same mixing matrix $\mathbf{M}$. It shows that at a critical error rate near 0.01 there is a sudden break in the graph and the oscillations abruptly appear. Below 0.01037 the system is stable with each weight vector of $\mathbf{W}$ converged at each of the independent components (ICs) whereas above 0.01037 oscillations suddenly appear. The change in behaviour at the critical error rate resembles a bifurcation from a stable fixed point, which represents a degraded version of the correct IC, to a limit cycle.

Different mixing matrices gave qualitatively similar results but the exact critical error value varied (see below). The results in Figs 2, 3 and 4A were obtained with $\gamma = 0.01$. Lowering the error rate produces very minor, and probably insignificant, changes in the critical error rate. Figure 4B shows the behavior at much lower learning rates (0.0005) for a different $\mathbf{M}$ (seed 10), over a long simulation period (150M epochs). The introduction of $b = 0.0088$ ($E = 0.0173$) at 4M epochs lead to a slow drop in the cosine which then crept down further until the sudden onset of a very slow oscillation at 35 M epochs; the next oscillation occurred at 140 M epochs. With $b = 0.00875$ ($E = 0.0172$) learning was perfectly stable over 68 M epochs, though degraded (data not shown). In this case the critical threshold appears to lie between 0.00875 and 0.0088, though possibly there are extremely slow oscillations even at 0.00875.

If γ was increased to 0.005 there was no clear change in the critical error rate. There was no oscillation within 60M epochs at 0.0086 error (using seed 10) but an oscillation appeared (after 4M epochs) at 0.0875 (see below). However the "oscillations" close to the critical error are quite irregular: at $b = 0.0088$ (γ = 0.005) the oscillation frequency was 4.18 +/- 0.31 mean +/-SD; range 4.41 to 3.64; $n = 5$); at 0.087 they were even slower (around 30M epochs) and more variable, and the weights changed in a steplike manner (see Text S1, section 2).

To explore the range of the critical error rate, 20 consecutive seeds (50-70) for **M**, i.e. 20 different random **M**s (with elements from {-1,1}), were used in simulations. One of the **M**s did not show oscillations at any error although two of the weights started to diverge without limit. The average critical per-connection error $b$ for the remaining 19 **M**s was 0.134, the standard deviation 0.16, the range 0.00875-0.475. In all these cases the critical error was less than the trivial value.

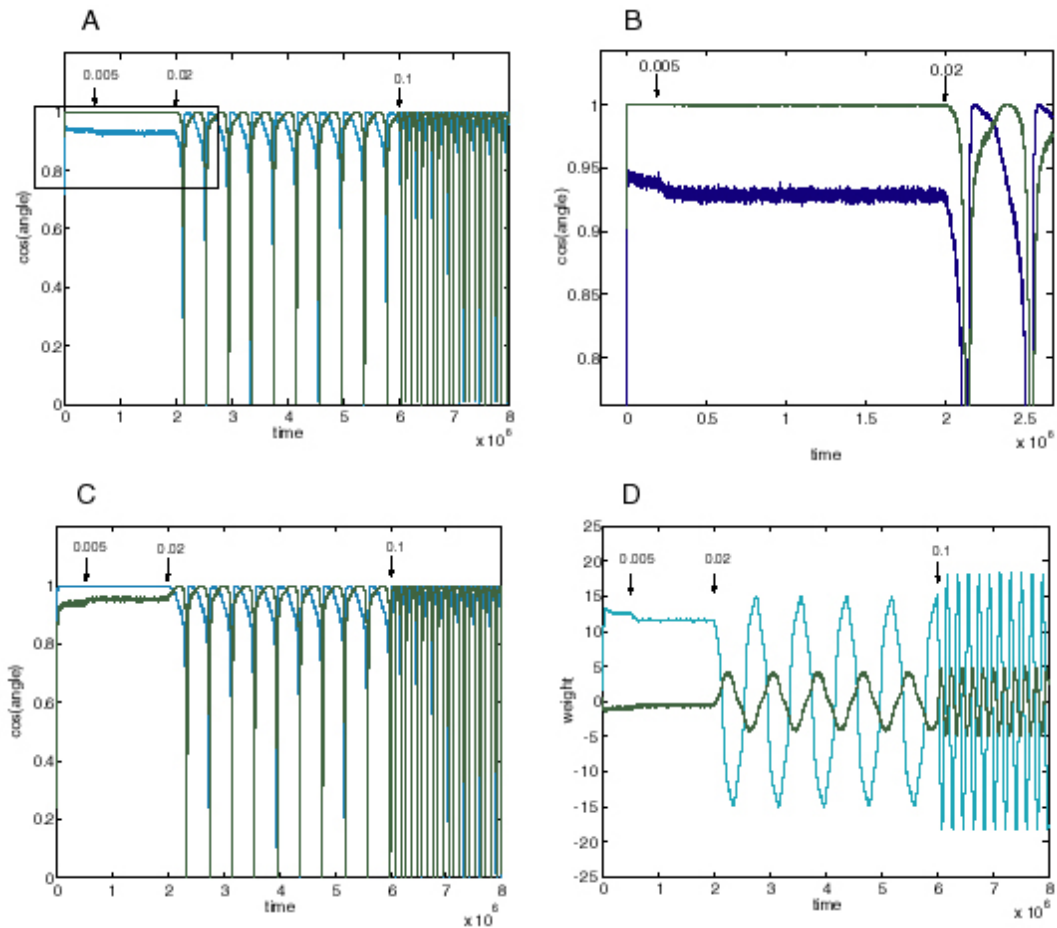

Figure 2  Plots 2A and 2C shows the initial convergence and subsequent behaviour, for the first and second rows of the weight matrix **W**, of a BS network with 2 input and 2 output neurons Error of $b = 0.005$

($E = 0.0099$) was applied at 200,000 epochs, $b= 0.02$ ($E= 0.0384$) at 2,000,000 epochs. At 6.000,000 epochs error of 0.1 ($E= 0.166$) was applied. The learning rate was 0.01. Figure 2A shows the first row of **W** compared against both rows of $\mathbf{M}^{-1}$ with the y-axis the cos(angle) between the vectors. In this case row 1 of **W** converged onto the second IC, i.e. the second row of $\mathbf{M}^{-1}$ (green line), while remaining at an angle to the other row (blue line). The weight vector stays very close to the IC even after error of 0.005 is applied, but after error of 0.02 is applied at 2,000,000 epochs the weight vector oscillates. Figure 2B is a blow-up of the box in Figure 2A showing the very fast initial convergence (vertical line at 0 time) to the IC (green line), the very small degradation produced at $b = 0.005$ (more clearly seen in the behavior of the blue line) and the cycling of the weight vector to each of the ICs that appeared at $b = 0.02$. It also shows more clearly that after the first spike the assignments of the weight vector to the 2 possible ICs interchanges. Figure 2C shows the second row of **W** converging on the first row of $\mathbf{M}^{-1}$, the first IC, and then showing similar behaviour. The frequency of oscillation increases as the error is further increased (0.1 at 6,000,000 epochs). Figure 2D plots the weights of the first row of **W** during the same simulation. At $b = 0.005$ the weights move away from their "correct" values, and at $b = 0.02$ almost sinusoidal oscillations appear.

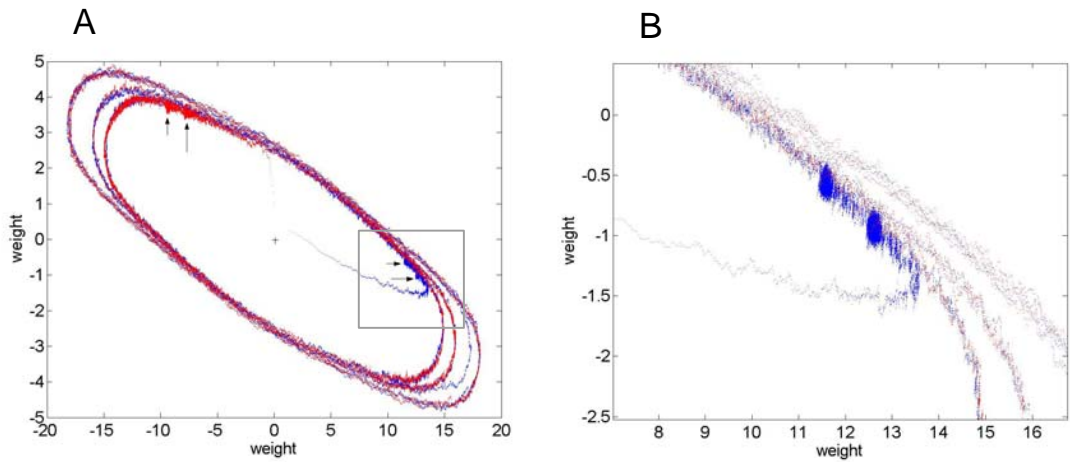

Figure 3 Trajectories of weights comprising the ICs. The weights comprising each IC (rows of the weight matrix) were plotted against each other over time (Figure 3A; red plot is the first row of **W** and the blue plot is the second row of **W**). The simulation was run for 1M epochs with no error applied and each row of **W** can be seen to evolve to an IC (red and blue 'blobs' indicated by large arrows in panel A). From 2M to 4M epochs error $b = 0.005$, i.e. below the critical error level, was applied and each row of **W** readjusts itself to a new stable point, red and blue 'blobs' indicated by the smaller arrows. From 4M to 6M epochs error of 0.02 was applied and each row of **W** now departs from a stable point and moves off onto a limit cycle-like trajectory (inner blue and red ellipses). Error is increased at 6M epochs to 0.05 and the trajectories are pushed out into longer ellipses. At 7M epochs error was increased again to 0.1 and the ellipses stretch out even more. Notice the transition from the middle ellipse to the outer one (error from 0.02 to 0.1) can be seen in the blue line (row 2 of **W**) in the bottom left of the plot. Figure 3B is a blow-up of the inset in Figure 3A clearly showing the stable fixed point of row 2 of **W** (i.e. an IC) at zero error (right hand blue 'blob'). The blob moves a small amount to the left and upwards when error of 0.005 is applied indicating that a new stable fixed point has been reached. Further increases in error launch the weights into orbit. $\gamma = 0.005$.

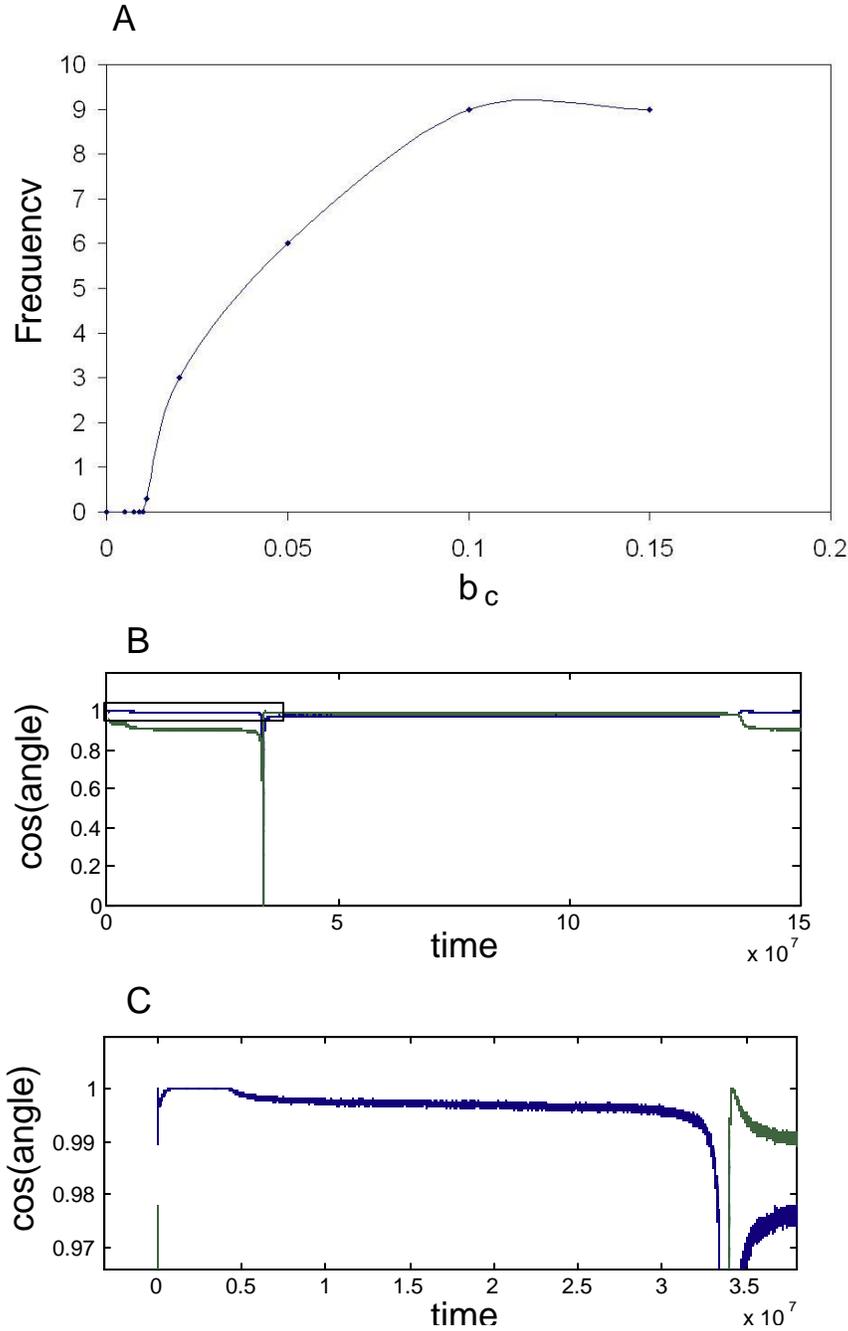

Figure 4. Figure 4A shows that increased error increases the frequency of the oscillations (cycles/$10^6$ epochs) but that the onset of oscillations is sudden at $b = 0.01$ ($E = 0.0196$; L=0.01 seed = 8), indicating that this critical error level heralds a new dynamical behaviour of the network. In B and C (enlargement of the box in B) the behaviour of the network at a very low learning rate is shown for a different learning rate and **M** ($\gamma = 0.0005$; seed = 10). The blue curves show cos(angle) with respect to the first row of $\mathbf{M}^{-1}$, the green curves with respect to the second column. Only the results for one of the output neurons is shown (the other neuron responded in mirror-image fashion). Plot B shows that the weight vector converged rapidly and precisely, in the absence of error, to the first row (blue curve; the initial convergence is better seen in C); error ($b = 0.0088$ $E = 0.0173$) was introduced after 5 million epochs; this led to a slow decline in performance over the next 5 million epochs to an almost stable level which was followed by a further

very slow decline over the next 30 million epochs (blue trace in C) which then initiated a further rapid decline in performance to zero (the downspike in B) which was very rapidly followed by a dramatic recovery to the level previously reached by the green assignment; meanwhile the green curve shows that the weight vector initially came to lie at an angle about $\cos^{-1} 0.95$ away from the second row of $\mathbf{M}^{-1}$.. The introduction of error caused it to move further away from this column (to an almost stable value about $\cos^{-1} 0.90$), but then to suddenly collapse to zero at almost the same time as the blue spike. Both curves collapse down to almost zero cosine, at times separated by about 10,000 epochs (not shown); at this time the weights themselves approach zero (not shown). The green curve very rapidly but transiently recovers to the level ($\cos(\theta) \sim 1$) initially reached by the blue curve, but then sinks back down to a level just below that reached by the blue curve during the 5M – 30M epoch period. Thus the assignments (blue to the first row initially, then green) rapidly change places during the spike by the weight vector going almost exactly orthogonal to *both* rows, a feat achieved because the weights shrink briefly almost to zero. During the long period preceding the return swap, one of the weights hovers near zero. After the first swapping (at 35M epochs) the assignments remain almost stable for 120M epochs, and then suddenly swap back again (at 140M epochs). This time the swap does not drive the shown weights to zero or orthogonal to both rows. However, simultaneous with this swap of the assignments of the first weight vector, the second weight vector undergoes its first spike to briefly attain quasi-orthogonality to both nonparallel rows, by weight vanishing (not shown). Conversely, during the spike shown here, the weight vector of the second neuron swapped its assignment in a nonspiking manner (not shown). Thus the introduction of a critical amount of error causes the onset of rapid swapping, although during almost all the time the performance is very close to that stably achieved at a just subcritical error rate ($b = 0.00875$; see Text S1, figure S1).

**Larger Networks**

Figure 5 shows a simulation of a network with $n = 5$. The behaviour with error is now more complicated. The dynamics of the convergence of one of the weight vectors to one of the rows of the correct unmixing matrix $\mathbf{M}^{-1}$ (i.e. to one of the five ICs) is shown (Figure 5A) using a random mixing matrix $\mathbf{M}$ (see Text S1, section 2). Figure 5A plots $\cos(\theta)$ for one of the five rows of $\mathbf{W}$ against one of the rows of $\mathbf{M}^{-1}$. An error of $b = 0.05$ ($E = 0.09$) was applied at 200,000 epochs, well after initial error-free convergence. The weight vector showed an apparently random movement thereafter, i.e. for 8 million epochs. Figure 5B shows the weight vector compared to the other rows of $\mathbf{M}^{-1}$ showing that no other IC was reached. Weight vector 2 (row 2 of $\mathbf{W}$) shows different behaviour after error is applied (Figure 5C). In this case the vector undergoes fairly regular oscillations, similar to the $n=2$ case. The oscillations persist for many epochs and then the vector (see pale blue line in Figure 5D) converged approximately onto another IC (in this case row 3 of $\mathbf{M}^{-1}$) and this arrangement was stable for several thousand epochs until oscillations appeared again, followed by another period of approximate convergence after 8.5 million epochs.

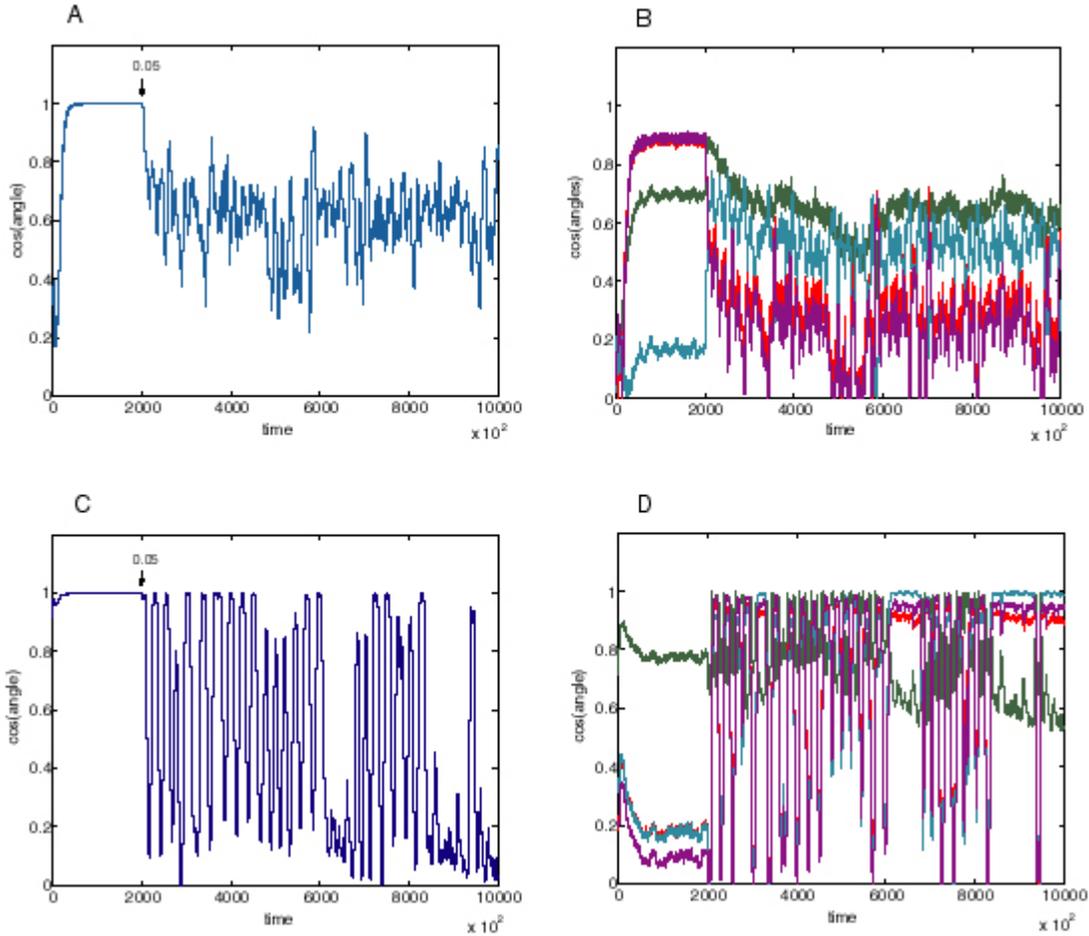

Figure 5        Figure 5A shows the of convergence of one of the rows of $\mathbf{M}^{-1}$, with one of the weight vectors of $\mathbf{M}$ (seed 8) with $n$=5. The initial weights of $\mathbf{W}$ are random. The angle between row 1 of the weight matrix and row 1 of the unmixing matrix are shown. The plot goes to 1 (i.e. parallel vectors) indicating that an IC has been reached. Without error this weight vector is stable. At 200,000 epochs error of 0.05 ($E$ = 0.09) is introduced and the weight vector then wanders in an apparently random manner. Figure 5B shows the weight vector compared to all the other potential ICs and clearly no IC is being reached. Plots 5C and 5D on the other hand shows different behaviour for row 2 of the weight matrix (which initially converged to row 4 of $\mathbf{M}^{-1}$). In this case the behaviour is oscillatory after error (0.05 at 200,000 epochs) is introduced, although another IC (in this case row 3 of $\mathbf{M}^{-1}$ (pale blue line) after 6.5M and again at 8.5M epochs) is sometimes reached, as can be seen in Figure 5D where the weight vector is plotted against all row of $\mathbf{M}^{-1}$. The learning rate was 0.01.

**Orthogonal mixing matrices**

ICA learning rules work better if the effective mixing matrix is orthogonal, so the mix vectors are pairwise uncorrelated (whitened). For $n$=2 we looked at the case where the data were whitened to varying extents. This could be done either by limiting the number of data vectors used to estimate $\mathbf{C}$, or by variably perturbing the whitening matrix $\mathbf{Z}$ (see Methods). We looked at the relationship between degree of perturbation from orthogonality of the whitened mixing matrix $\mathbf{Q} = \mathbf{ZC}$ and the onset of oscillation with error (see Methods). We found that there was a correlation (Figure 6, left graph) with the onset of oscillation occurring at lower error rates as **Q** was more and more perturbed from being orthogonal. Figure 6 (right graph) shows the effect of lowering the batch number used in estimating the covariance matrix **C** of the set of source vectors that have been mixed by a random matrix **M**. As the effective mixing matrix, which is orthogonal with perfect whitening, becomes less orthogonal (due to a cruder estimate of the decorrelating matrix by using a smaller batch number for the estimate of **C**) the onset of oscillations occur at lower and lower values of error.

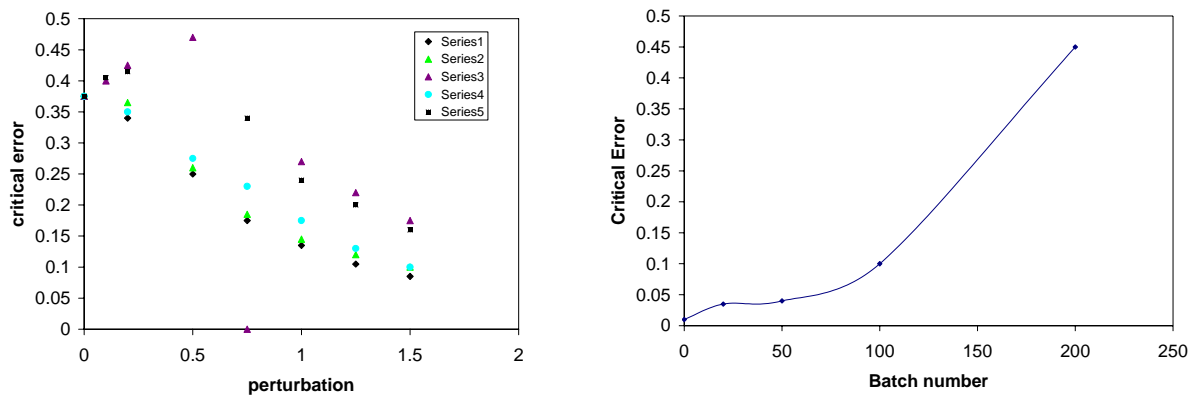

Figure 6 Relationship between degree of whitening with critical error. Figure 6 (left) shows the relationship between degree of perturbation of an orthogonal (whitened) matrix **Q** (seed 2, $n=2$) and the onset of oscillation. Data using five different perturbation matrices (series 1-5) applied to a decorrelating matrix **Z** (see Methods), are plotted. Each series is of one perturbation matrix, scaled by varying amounts (shown on the abscissa as "perturbation"), which is then added to **Z** (calculated from a sample of mixture vectors), and plotted against the critical error (obtained from running different simulations using each variably perturbed Z), shown on the ordinate. At zero perturbation the network still became unstable at a non-trivial error rate. As the orthogonal matrix was made less and less orthogonal by perturbing each of the elements of the decorrelating matrix **Z** (see Methods) by a small amount (0-25%) the sensitivity to error increased. The right hand graph is a plot for one random **M** ($n=5$, seed 8) where the mixed data has been whitened by a decorrelating matrix, $(\mathbf{C}^{\frac{1}{2}})^{-1}$. In this case the covariance matrix **C** of the mix vectors was estimated by using different batch numbers with a smaller batch number giving a cruder estimate of **C** which generates a less orthogonal mixing matrix. The learning rate was 0.01 in both graphs.

We noted above that the critical error rate for oscillation onset varies unpredictably for different **M**. There seemed to be no relationship between the angle between the columns of **M** and $b_c$ (not shown). In order to try to find a relationship between a property of a given random mixing matrix and the onset of oscillation, we plotted the ratio of the eigenvalues $\lambda_2$ and $\lambda_1$ of $\mathbf{MM}^T$ against $b_c$. If **M** is an orthogonal matrix then $\mathbf{MM}^T$ is the identity and $\lambda_2/\lambda_1$ is 1. If **M** is not orthogonal then the ratio is less than 1. We used the ratio $\lambda_2/\lambda_1$ as a measure of how orthogonal **M** was, and Figure 7 (left graph) shows a plot of this ratio against $b_c$ for the respective **M**. Although the points are scattered, there does appear to be a trend: as the ratio gets closer to 1, the value for $b_c$ gets larger. Figure 7 (right graph) is a plot of $\cos(\theta)$ of the normalized columns of the mixing matrices in Figure 7 (left graph) against the redetermined $b_c$ in runs for the adjusted **M**. There is a clear trend indicating that the more orthogonal the angle between the normalized columns of **M**, the less sensitive to error the matrix is. A few of the 'normalized' matrices, however, did not show oscillation at any value of error, perhaps because for some of these perturbed matrices the weights seemed to be growing without bound (there is no explicit normalization in the BS rule). The angles between the columns in these cases were always quite large. Completely orthogonal matrices were not, however, immune from sudden instability (i.e. at a critical error value $b_c$), as the two points lying on the x-axis in Figure 7 (right graph) demonstrate; here the angle between the columns is 90 degrees but there was a critical error rate at $b = 0.4$ and $0.45$, well below the trivial value.

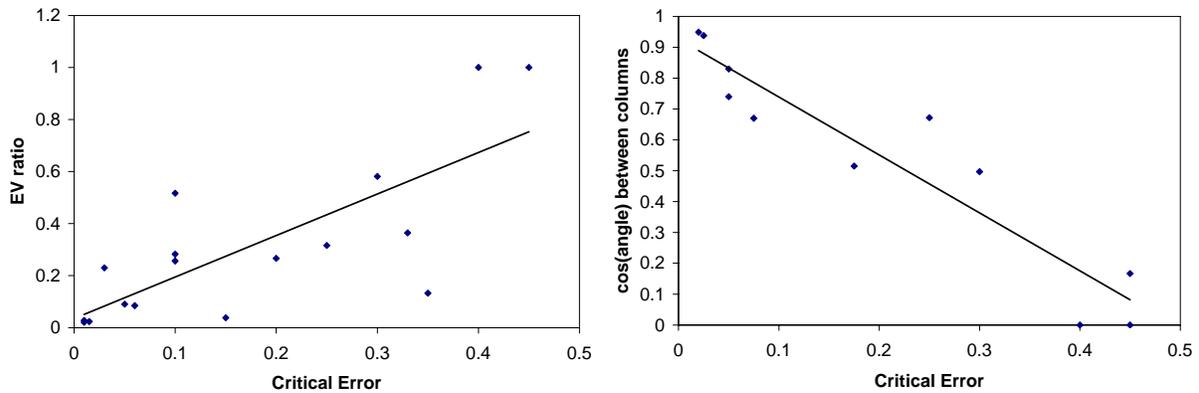

Figure 7    Relationship of increasing orthogonality of M with critical error. Figure 7 (left) shows a plot of the ratio of eigenvalues of $\mathbf{MM}^T$ ($\lambda_2/\lambda_1$) against the critical error $b_c$ for a given **M**, for various randomly-generated **M**s selected to give a range of critical errors. On the right hand side is a plot of the $b_c$ (Critical Error) against the cos(angle) between normalized columns of **M** (for the same set of random **M**s). Note that for 2 exactly orthogonal **M**s, different $b_c$ values were obtained. $n=2$. The lines in both graphs are least squares fits.

The source distribution was usually Laplacian, but some simulations were done with a logistic distribution (i.e. the distribution for which the nonlinearity is "matching"). The results were similar to those for the Laplacian distribution in terms of convergence to the ICs, but the onset of oscillation occurred at a critical error rate that was about half that for the Laplacian case, using the same random mixing matrices (data not shown).

**Discussion**

**Biological Background**

A synaptic connection has 2 main functions: it must convey selective information about the activity of the presynaptic neuron and its own current strength to the postsynaptic neuron, and it must appropriately adjust its strength based on the history of signals arriving at that connection. Both of these operations should occur independently at different connections, even though the individual synapses comprising connections are very small and densely packed. Optimizing these related but different functions must be quite difficult, especially since they are somewhat contradictory: some signals must pass through the synapse towards a spike trigger region, while other signals must be confined to the synapse itself. More specifically, electrical signals should spread, while chemical signals (extracellular transmitters and intracellular second messengers) should not. Compartmentalization is achieved by a combination of narrow extracellular and intracellular spaces and buffering/pumping mechanisms. However, these 2 compartmentalizing strategies are themselves contradictory: the chemicals that power pumps must arrive through the same narrow spaces. It is unlikely that connections operate completely independently of each other, even though there is little advantage in having large numbers of connections and neurons if they cannot. A central problem in neurobiology is the storage of information at very high density, as in other forms of computing (silicon or genetic), and neural information cannot be accurately stored unless connections change strength independently.

We are interested in the possibility that sophisticated brains use dual, direct and indirect, strategies to achieve high levels of connectional independence. Placing synapses on spines would be an example of a direct strategy. It is clear that the spine neck provides a significant, though not complete, barrier to calcium movement, and that calcium is a key chemical mediating activity-dependent modifications in synaptic strength [40-47,58]. We have proposed [48-51] that the neocortex might in addition use an indirect, "Hebbian proofreading", strategy, involving complex microcircuitry that independently monitors and regulates activity at connections. However, while the suggestions that synapses cannot operate completely independently, and that the neocortex is primarily a device for mitigating the effects of synaptic interdependence, are not inherently implausible, the key step in this argument has hitherto been missing: a demonstration that learning in realistic networks completely fails if synapses are not sufficiently independent.

**Physical Basis of Error and Nonlinearity**

Although in at least some cases coincident activity at one synapse does affect adjustments at others on the same neuron [11,12,15], the physical basis of such crosstalk is uncertain. In at least one case [52] crosstalk seems to be caused by dendritic diffusion of calcium. In a recent elegant study of crosstalk [15], evidence was obtained that crosstalk is caused by an "intracellular diffusible factor". However, these authors suggests that this factor was not calcium, since in their experiments the calcium increase at a synapse caused by an LTP-inducing protocol at a neighboring synapse was only 1% (and not significantly different from 0%) of that occurring at that neighboring synapse. However, this reasoning may be flawed. First, that 1% signal is even less significantly different from 1% than it is from zero, and could double the calcium concentration at that synapse. Second, the space constant for the dendritic diffusion of the "factor" was similar to that measured for calcium diffusion [40]. Third, immediately following an LTP-inducing protocol at a spiny synapse, there is a dramatic decrease in the diffusional coupling of the spine head to the shaft [53], which would presumably prevent the escape of any "factor" (except for calcium itself, which is the earliest spine head signal, and which presumably triggers the uncoupling). Fourth, since LTP at single synapse produces a stochastic, all-or-none increase in strength [54,55], and to reliably induce LTP adequate stimuli must be presented many times (e.g. 30 stimuli over 1 minute in the Harvey/Svoboda [15] experiments) it seems that some mechanism must "integrate" the magnitude of those stimuli over a minute-long time-window. An obvious "register" candidate for such integration is phosphorylation of CaM Kinase, the principal link between calcium and LTP expression [56-58]. This means that repeated small increases in calcium at a synapse that are in themselves insufficient to trigger LTP, could nevertheless be registered at that synapse, and add to subsequent subthreshold calcium signals at that synapse to trigger all-or-none LTP. In the reverse protocol [15], where the subthreshold remote stimulus is given first, no threshold change is seen, possibly because the observed spine structural changes shield the synapse from subsequent small dendritic calcium signals.

One possible objection to this argument would be that very small changes in calcium may fail to affect the register, for example if calcium activates CaM Kinase nonlinearly [59]. This raises the important question of the possible biophysical basis of the nonlinearity that is essential for learning high-order statistics. There are 2 possible limiting cases. (1) "nonlinearity first": the nonlinearity is applied to the Hebbian update *before* part of that update leaks to other synapses. This is the form we adopted in this paper (Eq (2)). In this case the nonlinearity might reflect a relation between depolarization and spiking, or between spike coincidence and calcium entry. (2) "nonlinearity last": the calcium signal would linearly relate to the number of coincidences; after attenuation it would then be linearly distributed to neighboring synapses, where it would nonlinearly combine with whatever other calcium signals occur at those synapses. This would lead to an equation of form:

$$\Delta \mathbf{W} = \gamma \ ([\mathbf{W}^T]^{-1} + [(\mathbf{1} - 2f(\mathbf{uE})\ \mathbf{x}^T])$$

We will describe the behavior of this case in another paper, but it seems to be similar to that described here.

Clearly in the "nonlinearity first" case, the register would respond linearly to calcium. In the "nonlinearity last" case, the register could perhaps discriminate against very small calcium signals emanating from neighboring synapses; however, the effectiveness of such a mechanism would be constrained by the requirement to implement a nonlinearity that is suitable for learning, and not just for discrimination against stray calcium. Furthermore, the finite background (resting) intracellular calcium concentration would complicate such discrimination.

None of our conclusions hinge on the nature of the diffusing crosstalk signal. However, if we assume it is calcium, we can try to estimate the magnitude of possible biological crosstalk, and compare this to our range of values of $b_c$. There are 2 possible approaches. The first is based on detailed realistic modeling of calcium diffusion along spine necks, including buffering and pumping. Although indirect, such modeling does not require the use of perturbing calcium-binding dyes. Zador and Koch [39] have estimated that about 5% of the calcium entering through the NMDAR might reach the dendritic shaft (most of the loss would be due to pumping by the spine neck membrane). How much of that 5% might reach neighboring spine heads? Obviously simple dilution of this calcium by the large shaft volume would greatly attenuate this calcium leakage signal, and then the diluted signal would be further attenuated by diffusion through a second spine neck. It might seem impossible that after passing this triple gauntlet (neck, dilution, neck) any calcium could survive. However, one must consider that the amount of stray calcium reaching a particular spine head reflects the combined contribution of stray signals from all neighboring spines: it will depend on the linear density of spines. One way to embody this was outlined in Methods. Another even simpler approach was adopted by Cornelisse et al [60]: they pointed out that in the case where all synapses are active together (perhaps a better approximation than that only one is active at a time) one could simply regard each spine as coupled to a shaft segment that was as long as the average distance between spines. Typically, this segment volume is comparable to the spine head volume, so the "dilution factor" would only be around 2-fold.

A second consideration is that the effect of neck pumps on calcium transfer from shaft to head will be much less than that on transfer from head to shaft, because the spine head does not have a large volume relative to the relevant dendritic segment. Indeed, the extra head-head attenuation produced by dilution is offset by the reduced head-head attenuation due to finite head volume. The underlying cause is the different boundary condition for head-shaft and shaft-head reaction-diffusion. To a first approximation, calcium movement along the neck can be modeled by a "calcium cable equation" which is identical in form to the electrical cable equation [39]. The reason for the 5% drop in the head shaft case is that the steady-state calcium concentration along the neck falls as sinh $(L-X)/L$ where L and X are position and neck length relative to the neck calcium "space constant"; the net flux to the shaft goes as $e^{-L}$ [61]. Recent data [62] show that the calcium pumps are strategically placed along the neck, as required for this analysis (calcium pumps in the spine head would be self-defeating: they also attenuate the calcium signal). To a first approximation, the spine head acts as a infinite calcium resistance, so the decay of calcium along the neck follows cosh $(L-X)/L$; if the neck is only 1 space constant long the expected attenuation is on the order 65%. Oddly enough, Zador and Koch illustrate a case of shaft-head diffusion in which the attenuation is 100%, possibly because they assumed the wrong boundary condition. The underlying reasoning here is exactly the same as that underlying the fact that electrical currents injected in dendrites produce voltages that fall much more rapidly towards the cell body than back towards the dendritic tips [63].

The second approach is direct measurement using fluorescent dyes. Such dyes inevitably perturb measurements, and this field has been very controversial, with one group claiming that under natural conditions there is negligible loss to the shaft [64] and other groups arguing that there can be low but significant loss [40, 61, 65]. We will present a detailed analysis elsewhere but on balance these studies suggest that natural loss is in the range 1-30%. A very conservative overall figure of 1% for head-shaft attenuation and 10% for shaft-head attenuation, giving a combined a value of $10^{-3}$, is used below.

**Error in the BS Model**

We studied the role of error in the BS model of ICA, an extensively studied learning paradigm in neural networks [22,23]. Figure 2 shows that the performance of the ICA network is at first only slightly degraded when minor error is introduced. It appears that the effect of minor crosstalk is that a slightly degraded version of $\mathbf{M}^{-1}$ is stably learned, as one might expect. This result is similar to what we see with linear Hebbian learning: the erroneous Oja rule [66] learns not the leading eigenvector of the input covariance matrix $\mathbf{C}$, but that of $\mathbf{EC}$ ([67]; our unpublished work with A. Radulescu). However, in the linear case, stable (though degraded) learning occurs all the way up to the trivial error rate. It appears that in the present, nonlinear, case, at a critical error rate a qualitatively new behaviour emerges: weight vectors begin to move in an oscillatory manner between approximately correct solutions.

To some extent this could be viewed as a manifestation of the freedom of the BS rule to pick any of the possible permutations of $\mathbf{M}^{-1}$ that allow source recovery, and if we had measured performance using the Amari distance [68], which takes into account all possible assignments, the oscillations would be concealed. An extreme case would be if the weights instantaneously jumped between correct assignments, as seems to happen exactly at the error threshold (see Text S1, section 2): this would not affect the Amari distance and within the strict ICA framework, any $\mathbf{W}$ that allows sources to be correctly estimated is valid. Such jumps are usually never seen in the absence of error, and to our knowledge such behavior has never been reported (though we have observed approximately this behavior in error-free simulations using high learning rates, which are of course very noisy). "Jumping", or "tunneling", is never normally seen because it can only occur (at low learning rates) if weights shrink to zero, but it appears that the addition of crosstalk errors allows such jumping. At very low learning rates and very close to the critical error rate, one weight hovers near zero, allowing "jumps" (i.e. the spike-like weight vector adjustments in Figure 4B and Text S1, section 2). Then the weight vectors rapidly adopt new assignments, and the Amari distance would be an almost equal quantity on either side of the error threshold. Furthermore, since the interspike intervals get increasingly irregular as error is reduced toward $b_c$, these jumps are essentially spontaneous: one cannot predict when they occur (though they presumably depend on the detailed historical sequence of the source vectors). At higher learning rates (Figure 2) or for error rates well beyond $b_c$ (Figure 4A), the network spends relatively more time relearning a progressively less accurate permuted version of $\mathbf{M}^{-1}$, so the Amari distance (averaged over many epochs) would decline further.

However, even if the inaccurate network jumps between more-or-less "useful" assignments, we suggest that it would be biologically useless, because in the brain ICA-like networks are used for further processing, and learning in those downstream networks presupposes stability of upstream networks. Thus, we (and others: [24,25]) suggest that even though the complex real-world process that generates input to the brain is unlikely to be simple linear mixing, a reasonable, ICA-like, early learning strategy, in the neocortex, would be to assume that they are so generated. Similarly, an *approximate* subcortical strategy in the initial stages of learning would be PCA-like (removal of pairwise correlations), since this is the best strategy when signals are Gaussian. The fact that neither statistical model is obeyed exactly does not mean that they are not good first guesses: after all, provided little information is *discarded* as a result of such preprocessing, the new representations can be more useful than the old. In particular, after maximizing mutual information under the assumption of an approximately linear mixing model, the components of the new representation (the *u* or *y* values in the present case) have residual even higher order correlations that can still, in principle, be learned from. Of course to the extent that the assumed linear mixing is incorrect, the representation is suboptimal (and therefore more likely to be contaminated by "wetware" noise). However learning to exactly invert the true generative model is likely to be prohibitively difficult even in a hierarchical system. Of course the real brain does not have the luxuries of learning infinitely accurately, so instantaneous jumping between almost perfect assignments will be impossible, and the situation for downstream networks even worse. For all practical purposes, a front-end ICA-like network whose weights oscillate in the way we describe here would be useless, and therefore if such processes do occur in the brain (particularly in the cortex), the error-problem must have somehow been solved.

**The Dynamical Behaviour**

The behaviour seen beyond the critical error rate may arise because the fixed point of the dynamics of the modified BS rule, i.e. a degraded estimate of $\mathbf{M}^{-1}$, becomes unstable. The behavior in Figures 2, 3 and 4A resembles a bifurcation from a stable fixed point to a limit cycle, the foci of which correspond approximately to permutations of $\mathbf{M}^{-1}$. Although we suspect that this is the case, we have not yet proved it, since it is difficult to write an explicit expression for the equilibria of the erroneous rule. . Our unpublished work with Kim et. al. with other versions of the erroneous learning rule suggests that the real part of the eigenvalues of the Jacobian of the linearised erroneous learning equation goes positive at a critical error. If the system's Jacobian eigenvalues become complex and then cross the imaginary axis in the complex plane, the ensuing onset of instability can occur as either a subcritical or supercritical Hopf bifurcation [69]. In the latter case, a stable limit cycle emerges. The cases where oscillations are not seen, and the weights diverge as error is increased beyond a critical point, might correspond to a subcritical bifurcation. Figure 5A and 5B shows that when *n*=5 more complex behaviour can occur for error beyond the critical level. We see that one of the rows of **W** seems to wander irregularly, not visiting any IC for millions of epochs. We do not know if this behavior reflects a complicated limit cycle or chaos, but from a practical point of view this outcome would be catastrophic. In unpublished work we have also studied the effect of crosstalk in a slightly different ICA network, using a normalized 1-unit rule [31]. We found that when all the inputs but one are Gaussian (so that they are not candidate IC's for the learning process), while oscillations do not occur, learning still breaks down at a critical error rate. We suggest therefore that breakdown may be a generic feature of inaccurate learning in ICA, and possibly in all nonlinear learning.

**Whitening**

Most practical ICA algorithms use whitening (removal of pairwise correlations) and sphering (equalizing the signal variances) as preprocessing steps. In some cases the algorithms *require* that **M** be orthogonal (so the mixed signals are pairwise uncorrelated). As noted above it is likely that the brain also preprocesses data sent to the cortex (e.g. decorrelation in the retina and perhaps thalamus [70-72], and we explored how this would affect the performance of the inaccurate ICA network. Whitening the data did indeed make the network more robust to Hebbian error as Figures 6 and 7 show, with the onset of instability occurring at higher error levels as the data were whitened more. However, even for completely orthogonal **M**s, oscillations usually still appear at error rates below the "trivial value" $\varepsilon_s$, for which learning is completely inspecific ($\varepsilon_s = (n-1)/n$). As discussed further below, if synapses are too densely packed, even error rates close to the trivial rate could occur in the brain.

Neither for random nor orthogonal **M**s could we predict exactly where the critical error would lie, although it is typically higher in the orthogonal case (Figures 6, 7). Some, but not all, of the variation in the $b_c$ values could be explained by the degree of nonorthogonality of **M**, estimated in 2 different ways. First, for an orthogonal matrix multiplication by its transpose yields the identity matrix, which has all its eigenvalues equal; we found that the $b_c$ for a given random **M** was correlated with the ratio of the first 2 eigenvalues of **MM**$^T$ (Figure 7, left). Second, if the columns of a matrix whose columns are orthogonal have equal length (i.e. the matrix is orthogonal), so do the rows. When we normalized the columns of a given random **M**, we found an improved correlation between the cosine of the angle between the columns and $b_c$ (Figure 7). We are currently studying the interesting issue of the distribution of $b_c$ values and their relation to properties of **M.**

Another factor influencing the critical error rate for a given **M** was the source distribution; we found that the critical error rate was typically about halved for logistic sources compared to Laplacian, despite the fact that this improves the match between the nonlinearity and the source cdf. We suspect that this is because the kurtosis is lower for the logistic distribution (1.2 compared to 3 for the Laplacian).

Even though learning can tolerate low amounts of error in favorable cases (particular instances of **M** and/or source distributions), low error can only be guaranteed by using small numbers of inputs. In the neocortex the number of feedforward inputs that potentially synapse on a neuron in a cortical column often exceeds 1000 [38], so *b* values would have to be well below $10^{-3}$ to keep total error below the trivial value, and considerably less to allow learning in the majority of cases. In the simple model summarized in the Methods, which assumes that strengthening is proportional to calcium, which diffuses along dendrites, we obtained $b = 2 \alpha a \lambda_c / L$. $a$ is the effective calcium attenuation from one spine head to another when both are at the same dendritic location; a factor that the preceding discussion suggests cannot be much below $10^{-3}$. Alpha is typically around 10 for feedforward connections [38], $\lambda_c$ around 3 μm [40] and *L* around 1000 μm [38], so *nb* would be around $6 \times 10^{-2}$, which often produces breakdown for Laplacian sources. If the cortex is to learn reliably, it would require additional machinery, especially if input statistics were less rich in higher order correlations than in our Laplacian simulations (see below).

The fact that whitening can make the learning rule almost error immune suggests at first sight that our study has only theoretical, not practical, significance, because whitening is a standard process which digital computers can accurately implement. However, the brain is an analog computer (albeit massively parallel) and so it cannot whiten perfectly, because whitening filters cannot be perfected by inaccurate learning. While learning crosstalk does not produce a qualitative change in the performance of the Oja model of principal components analysis (unlike the ICA model studied here), it does degrade it, especially when patterns are correlated (our unpublished work with A. Radulescu). This means that learned whitening is not a panacea for crosstalk, though if done fairly accurately it helps. (Of course, if the whitening can be done using gene-based evolution to "learn" the whitening matrix, the impact of Hebbian crosstalk would be lessened, but such "learning" must also be imperfect since the size of the genome, and thus the accuracy of gene-based wiring, is limited by the accuracy of DNA replication; [17-20]). One might wonder whether the fact that exact whitening requires exactly linear learning, and biological implementations of learning rules cannot be exactly linear, might also impair neural whitening; however, it seems likely that inevitable albeit small amount of error would help linearize the learning rule).

**Learning in the Neocortex**

If the neocortex learns to do ICA, or something similar, our results suggest 2 possibilities. One is that the relevant cortical neurons would have very few inputs (e.g. from thalamus) so that the corresponding feedforward synapses could be placed far enough apart on the dendrites to minimize crosstalk. There are indeed indications that thalamic inputs are widely spaced on the dendrites of spiny stellate cells in barrel cortex [73]; the large numbers of intervening synapses are nonthalamic, and while *their* learning might also produce undesirable calcium spillover, this should not have high-order correlations with the pattern of feedforward input activity. However, such a "brute-force" solution (and others like it) suffers the drawback that the distance between synapses is constrained by the electrical properties of dendrites. More importantly, it seems that cortical cells often do have hundreds or even thousands of potential feedforward inputs [38], so that nonlinear learning is presumably done despite the inevitable synaptic crowding, and the subtlety and variability of the world. How can neocortical neurons learn from arbitrary higher-order correlations between large numbers of inputs even though their learning rules are not completely synapse-specific?

The root of the problem is that the spike coincidence-based mechanism which underlies linear or nonlinear Hebbian learning is not completely accurate: coincidences at neighboring synapses affect the outcome ("type 2" errors). (Another type of error, lack of precision, would arise if the size of the update were not exactly determined by the learning rule, as a result of stochastic effects; we have also studied these "type 1" errors, but here brute force solutions appear adequate to avoid breakdown). Other than self-defeating brute force solutions (e.g. narrowing the spine neck), the only way to handle such inaccuracy is to make a second independent measure of coincidence, and it is interesting that much of the otherwise mysterious circuitry of the neocortex seems well-suited to such a strategy. If 2 *independent* though not completely accurate measures of spike coincidence at a particular neural connection (one based on the NMDAR receptors located at the component synapses, and another performed by dedicated specialized "Hebbian neurons" which receive copies of the spikes arriving, pre- and/or postsynaptically, at that connection) are available, they can be combined to obtain an improved estimate of coincidence, a "proofreading" strategy [51]. The confirmatory output of the coincidence-detecting Hebbian neuron would have to be somehow applied to the synapses comprising the relevant connection, such that the second coincidence signal would allow the first (synaptic) coincidence signal to actually lead to a strength change. While direct application (via a dedicated modulatory "third wire") seems impossible, an effective indirect strategy would be to apply the proofreading signal globally, via 2 branches, to all the synapses made by the input cell and received by the output cell; the only synapses that would receive both, required, branches of the confirmatory feedback would be those comprising the relevant connection (in a sufficiently sparsely active and sparsely connected network; [74]).

**Implementing Neocortical "Proofreading"**

We have suggested that layer 6 neurons are uniquely suited to such a Hebbian proofreading role, since they have the right sets of feedforward and feedback connections [48,49]. In particular, feedback to thalamus could, via the reticular nucleus, disynaptically hyperpolarize selected sets of relay cells, allowing burst firing, which could enable plasticity in mature thalamocortical synapses, while disenabling plasticity in others by monosynaptic depolarization [75]. Also, layer 6 modulatory feedback to layer 4 dendritic shafts via drumstick synapses could enable a postsynaptic component of plasticity, perhaps via metabotropic glutamate receptors, [75]. The main objection to such a proofreading scheme is that at first glance it appears to require dedicated proofreading neurons for every synapse. However, this requirement could be avoided by (1) using multisynapse feedforward connections (2) only using proofreading for existing (not incipient or potential; [49,37]) connections; and (3) using the same proofreading neuron for *all* the presynaptic neurons that feedforward onto a given postsynaptic neuron (or vice versa); this achieves "distributed" rather than "dedicated" proofreading but should work well providing that neural activity is sparse. Fortunately, ICA learning generates sparse codes [76].

In summary, we have shown that if the nonlinear Hebbian rule that underlies neural ICA is insufficiently accurate, learning collapses. Since the neocortex is probably specialized to learn higher-order correlations using nonlinear Hebbian rules, a major task might be the mitigation of inevitable errors, by a type of Hebbian proofreading mechanism. The evolution of such machinery would enable the onset of sophisticated learning, understanding and "mind".

## Acknowledgments

We thank Larry Abbott and Terry Elliott for their comments on the manuscript, and to Miguel Maravall for discussions and input on an earlier draft.

# Hebbian Crosstalk Prevents Nonlinear Unsupervised Learning

*Supplementary Text (Text S1)*

**Section 1**

**Methods**

<u>Generation of random vectors</u>

To get a vector of which each element is drawn from a Laplacian distribution, first an N element vector **s**, the elements of which is drawn from a uniform distribution (range{-0.5,0.5}), is generated by using the Matlab rand function: **s** = -0.5 + (0.5-(-0.5)) * rand(1,N). Then each element $s_i$ of x is then transformed into a Laplacian by the following operation;
$s_i = - \text{sign}(s_i)\ln(1 - 2|s_i|)$

'sign' means take the variable $x_i$ and if it is positive, assign it the value 1, if it is negative assign it the value -1, and if 0 assign it the value 0.

<u>Mixing matrices used in the simulations</u>

The mixing matrix **M** used for figure 2 was $\begin{pmatrix} 0.034 & 0.128 \\ 0.455 & 0.281 \end{pmatrix}$ (rand seed 8, {0,1})

and for figure 4 was $\begin{pmatrix} 0.45 & 0.128 \\ 0.208 & 0.076 \end{pmatrix}$ (rand seed 10, {0,1})

The mixing matrix (seed 8) used in figure 5 was

$$\mathbf{M} = \begin{pmatrix} 0.03 & 0.25 & 0.67 & 0.26 & 0.84 \\ 0.45 & 0.60 & 0.15 & 0.23 & 0.20 \\ 0.12 & 0.88 & 0.87 & 0.78 & 0.95 \\ 0.28 & 0.96 & 0.001 & 0.94 & 0.44 \\ 0.99 & 0.75 & 0.91 & 0.72 & 0.35 \end{pmatrix}$$

<u>Orthogonality</u>

Perturbations from orthogonality were introduced by adding a scaled matrix (**R**) of numbers (drawn randomly from a gaussian distribution) to the whitening matrix **Z**. The scaling factor (which we call 'perturbation') was used as a variable for making **Q** less orthogonal, as in figure 5. Below are the matrices used in generating one of the data sets of figure 5 with **M** generated from seed 8 (seeds 2-6 were used to generate the different **R** matrices for the 5 data sets in figure 5):

$$\mathbf{Z} = \begin{pmatrix} 10.6 & -1.79 \\ -1.59 & 1.94 \end{pmatrix} \quad \mathbf{R} = \begin{pmatrix} 0.37 & -0.18 \\ -0.22 & 0.176 \end{pmatrix} \text{ (from seed 2)}$$

For instance the matrix at pert = 0.5 on the graph would be (0.5**R**+**Z**)**M**

## Section 2

## Results

Plots near the error threshold

Figure 4B showed a 150M epoch simulation using seed 10 for **M** b =0.0088 and $\gamma = 0.0005$. During the oscillation "spikes" one of the weight vectors moves almost exactly orthogonal to both of the rows of $\mathbf{M}^{-1}$. This can only happen if both weights go through zero at the same moment. Closer inspection revealed however that there is a slight delay (on the order of 10K epochs) between the moments that these vectors swing through 90 degrees, such that the 2 weights do not zero at exactly the same moment. Preceding the swings, one of the weights spends very long periods hovering near zero. At these very low learning rates, the weight vector spends extremely small amounts of time near any of the rows of $\mathbf{M}^{-1}$.

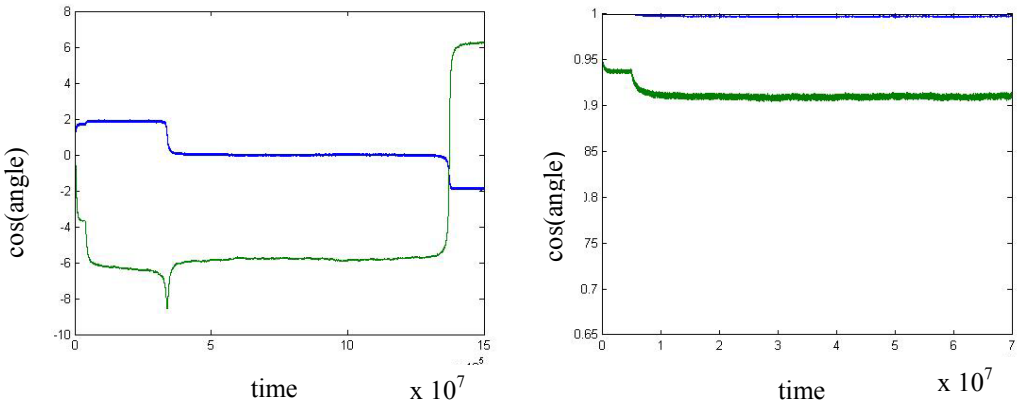

Figure S 1  On the left is a plot of the weights of one of the rows of **W** with error of 0.0088 (i.e. just above the apparent critical error) applied at 4M epochs at $\gamma = 0.0005$ (seed 10). These are the weights comprising the "other" weight vector from the one whose behavior was shown in Figure 4B and 4C. Thus the large swing in the weight vector shown in Fig 4B,C produced relatively small adjustments in the weights shown here ( at 30M epochs), while the very large weight changes shown here (at 140M epochs)correspond to small shifts in the direction of the weight vector shown in Fig 4B,C.(Conversely, these large weight steps at 140M epochs produce a spike-like swing in the corresponding weight vector angle). Note the weights make rapid steps between their quasistable values. Also the smaller (blue) weight spends a very long time close to zero preceding the large weight swing (during which swing the weight vector goes briefly and almost simultaneously orthogonal to both rows of $\mathbf{M^{-1}}$). Close inspection revealed that the blue weight crosses and recrosses zero several times during the long "incubation" period near zero. Note the wobbly appearance of the green weight. The thickness of the lines in the left and right plots reflects rapid small fluctuations in the weights that are due to the finite learning rate.

On the right is the plot of the cos(angle) between the weight vector whose components are shown in the left plot, and the 2 rows of $\mathbf{M^{-1}}$. However, b = 0.00875 (i.e. very close to the error threshold; see figure S2) introduced at 5M epochs; other parameters the same as in the left plot. Note that the weight vector relaxes from the correct IC to a new stable position corresponding to a cos angle just below 1 (blue plot), and then stays there for 65M epochs. The relaxation is more clearly seen in the green plot, which shows the cos angle with the row of $\mathbf{M^{-1}}$ that was not selected.

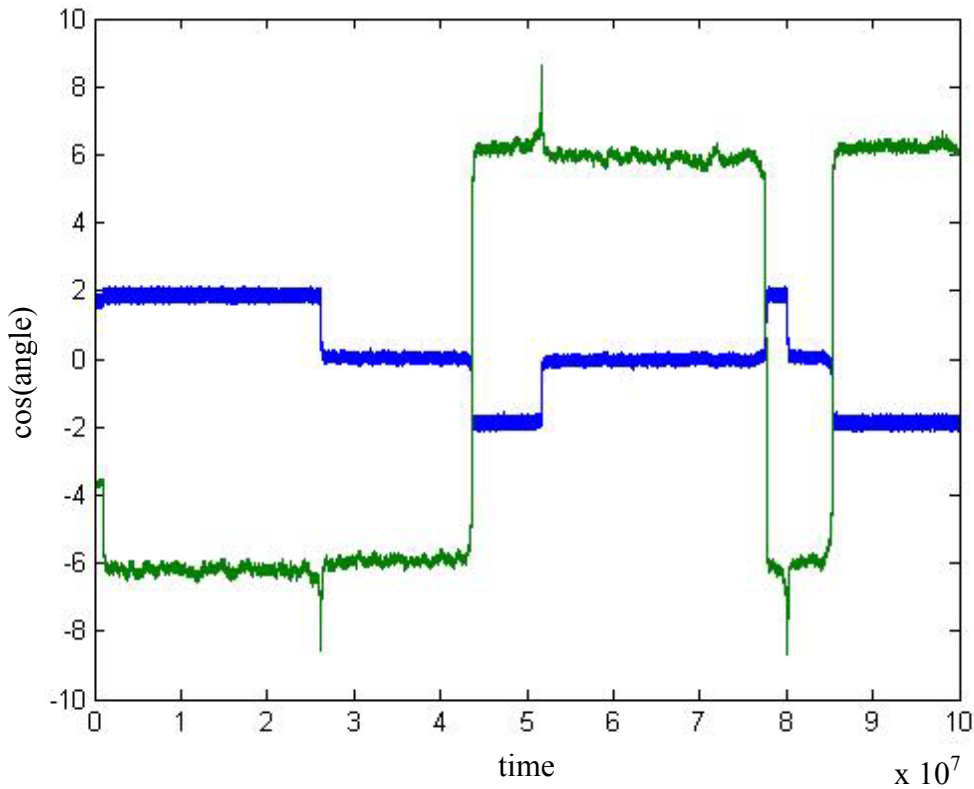

Figure S2. Plots of individual rates using the same parameters as in Figure S1 except $\gamma = 0.005$ (which increases the size of the slow and fast fluctuations, which is why the lines are thicker than in Fig S1) and b = 0.0087 (which appears to be extremely close to the true error threshold; the first oscillations occurs at 27M epochs, which would correspond to 270M epochs at the learning rate used in Figure S1), introduced at 1M epochs. . Each weight (i.e. green and blue lines) comprising the weight vector adopts 4 possible values, and when the weights step between their possible values they do so synchronously and in a particular sequence (though at unpredictable times). The 4 values of each weight occur as opposite pairs. Thus the green weight occurs as one of 4 large values, 2 positive and 2 equal, but negative. The 2 possible positive weights are separated by a small amount, as are the 2 possible negative weights. The blue weight can also occupy 4 different, but smaller values. Thus there are 2 small, equal but reversed sign weights, and 2 even smaller equal but reversed sign weights. These very small weights lie very close to zero. Since the weights jump almost synchronously between their 4 possible values, the "orbit" is very close to a parallelogram,

which rounds into an ellipse as error increases. One can interpret the 4 corners of the parallelogram as the 4 possible ICs that the weights can adopt: the 2 ICs that they actually do adopt initially and the 2 reversed sign ICs that they could have adopted (if the initial weights had reversed sign). However, 2 of the corners are closer to correct solutions than are the others (corresponding to the assignment reached when the blue weights are very close to zero). It seems likely that exactly at the error threshold the difference between the 2 close values of the green weights, and the difference between the very small values of the blue weights, would vanish. This would mean that the blue weights would be extremely close to zero during the long period preceding an assignment swap, so the direction of the weight vector would be very sensitive to the details of the arriving patterns. Consistent with this interpretation, the weights fluctuate slowly during the long periods preceding swaps; these fluctuations, combined with the vanishing size of one of the weights, presumably make the system sensitive to rare but special sequences of input patterns. Similar behavior was seen using seed 8.

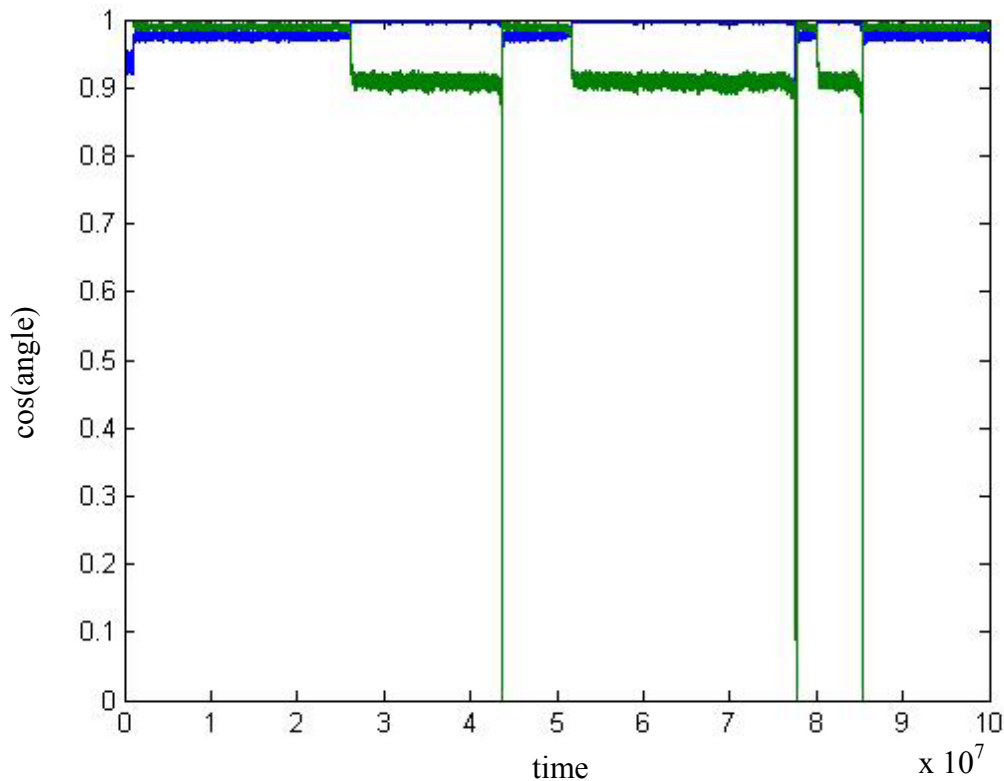

Figure S3 This shows the behavior of the weight vector whose component weights are shown in Fig S2 (cos angle with respect to the 2 rows of $\mathbf{M}^{-1}$) Error b = 0.0087 introduced at 1M epochs. Note the weight vector steps almost instantaneously between its 2 possible assignments. However, when the weight vector is at the blue assignment, it is closer to a true IC than it is when it is at the green assignment (which is the assignment it initially adopts. When the weight vector shifts back to its original assignment (at 43M epochs), it shifts orthogonal to both ICs at almost the same moment (sharp downspikes to zero cosine). Notice the extreme irregularity of the "oscillations".

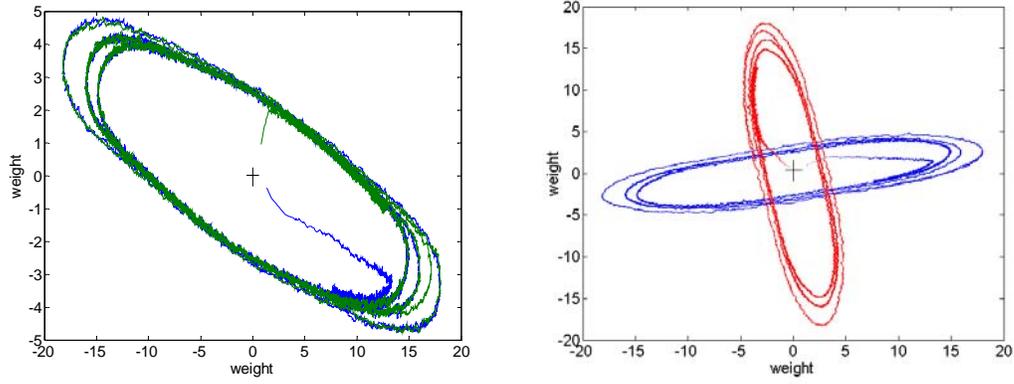

Figure S4  The plot on the right is similar to those of figure 3 except that the data was generated from a different simulation with all parameters being the same except that the initial weight vectors were different. Notice how one of the weight vectors (rows of **W**) initially evolves to the mirror image in terms of sign of the weight vector in Figure 3A (right most red blob). The right hand plot shows weight 1 from row 1 of **W** with weight 2 of row 2 (blue) and weight 2 of row 1 with weight 1 of row 2 (red).